# High-Resolution Coherent DFS Over 20km Ultra-Low-Loss Anti-Resonant Hollow-Core Fiber with Live Traffic

**Rajiv Boddeda[1], Arnaud Dupas[1], Haïk Mardoyan[1], Christian Dorize[1], Fabien Boitier[1], Peng Li[2], Zhang Lei[2], Jie Luo[2], Pierre Brochard[3], Carina Castineiras[1], Jelena Pesic[4], Florian Pulka[4] and Jérémie Renaudier[1]**

*(1) Nokia Bell Labs, 12 rue Jean Bart, 91300 Massy, France (2) State Key Laboratory of Optical Fiber and Cable Manufacture Technology, Yangtze Optical Fibre and Cable Joint Stock Limited Company (YOFC), Wuhan, Hubei, China. (3) Silentsys, 10 rue Xavier Bichat, 72000, Le Mans, France (4) Nokia Networks France, Optics division, Route de Villejust, 91620 Nozay,. [rajiv.boddeda@nokia-bell-labs.com](mailto:rajiv.boddeda@nokia-bell-labs.com)*

**Abstract:** We demonstrate sub-meter resolution Coherent DFS and detect acoustic oscillations using a stabilized laser on 20 km of anti-resonant HCF with <0.10 dB/km loss without impacting live traffic of 1.2 Tbps on the adjacent channel. © 2024 The Author(s)

## 1. Introduction

One of the most promising recent evolutions in optical telecom systems is the development of low-loss anti-resonant hollow core fibers (AR-HCF) with < 0.11 dB/km [1]. Unlike traditional solid core fibers, HCFs guide light through an air-filled core by significantly reducing the interaction between light and matter leading to lower latency, reduced attenuation while enabling high-power capable transmission systems. However, this reduced light-matter interaction leads to a lower backscattered power. In such fibers, Rayleigh Backscattering Coefficient (RBC) can be categorized into three primary types. The most significant contribution originates from air, with a backscattering level ranging from -90 to -100 dB/m. This is followed by backscattering from microstructure surfaces, which occurs at around -115 dB/m. Lastly, the glass regions of the fiber contribute to backscattering at an approximate level of -150 dB/m [2]. Hence, detecting any acoustic vibrations over HCF is highly challenging as the backscattering from glass regions is significantly low. In addition, current hollow core fibers are equipped with optical adapters to couple light between single mode fibers (SMF) and HCF fibers leading to strong parasitic reflections. Recently, the use of Distributed Acoustic Systems (DAS) was reported in [3] over photonic bandgap HCF with RBC around -92 dB/m with a propagation loss of about 2.5 dB/km. Chirped pulse DAS operation was demonstrated over 6.7 km however the coupling angle of the SMF/HCF adapter is increased to reduce the parasitic reflections and due to high propagation losses it leads to reduced reflection peak from the farther end. However, such a technique inherently induces higher losses which is not suitable for optical transport, particularly in the case of low loss HCF aiming to minimize the loss of optical transmission lines.

In this article, we demonstrate highly sensitive Distributed Fiber Sensing (DFS) with sub-meter resolution over 20.2 km of novel ultra-low loss support tube hollow core fiber (ST-HCF). First, we show the characterization of the 20.2 km of ST-HCF using a standard Optical Time Domain Reflectometers (OTDR) measurement and detail the limitations of the system. Achieving high resolution is challenging due to parasitic reflections originating from the adapter (approximately a few centimeters), necessitating measurement capabilities with a very high dynamic range. We overcome this limitation using our DFS system with stabilized laser to obtain sub-meter resolution. Moreover, we pinpoint precisely the splicing losses in the 20 km HCF consisting of two spools of continuous HCFs. Finally, we perform acoustic sensing by exciting one of the exposed regions of the fiber and successfully identify the excitation while maintaining uninterrupted live traffic at 1.2 Tb/s on an adjacent channel at launch powers up to 23 dBm.

## 2. Experimental setup

In this experiment, we utilize a novel type of ST-HCF developed with support columns, where the middle and large capillaries are separated to create a gap. This design increases the number of anti-resonant layers and suppresses the coupling between the core mode and the tube mode. These hollow core fibers, manufactured by YOFC, can achieve loss below 0.1 dB/km @ 1550nm. In Figure 1a, we show the characteristics of the HCF fiber consisting of two short SSMF fibers at the beginning and the end of the fiber coupled to the HCF using a mode field adapter (MFA). The mode field is effectively matched with the custom-designed graded-index fiber, and Fresnel reflection is minimized through angle-cutting and the application of an anti-reflection coating.

We perform OTDR using a commercial of the shelf device to obtain detailed information about the fiber's length, attenuation, and the presence of faults or splices. The pulse duration in OTDR is a critical parameter that significantly influences the resolution and the range of the measurements. Shorter pulse durations allow for higher resolution, enabling the detection of closely spaced events such as splices and faults. However, shorter pulses also reduce the measurement dynamic range due to lower energy levels especially for HCF fibers where the Rayleigh reflection is weak [2]. To the best of our knowledge, the highest resolved OTDR measurement is 1.5 m obtained using long acquisition time (~ hours) or resorting to pulse amplification techniques with undesirable effects [4]. In figure 1b, we show the OTDR measurement using a pulse length of 100 ns over 3 minutes of averaging leading to a Full Width at Half Maximum (FWHM) of around 30 m around spliced regions. Moreover, we observe a large dead zone area of around 1 km around the spliced regions. We observe strong Fresnel reflections coming from the MFA at both ends of the fiber limiting the dynamic range. In Figure 1b, we show that the RBC from the SMF patch (typical value of -76 dB/m) is about 18.1 dB higher than the extrapolated value at the beginning of the HCF. Hence, an RBC due to the air inside the HCF is estimated to be about -93.7 dB/m, which is consistent with reported values [5,6]. By performing OTDR in both directions, we estimate the propagation loss as a function of distance and a linear fit provides a value of 0.096 dB/km with an error of +/- 0.005 dB/km depending on the region of interest for the fit.

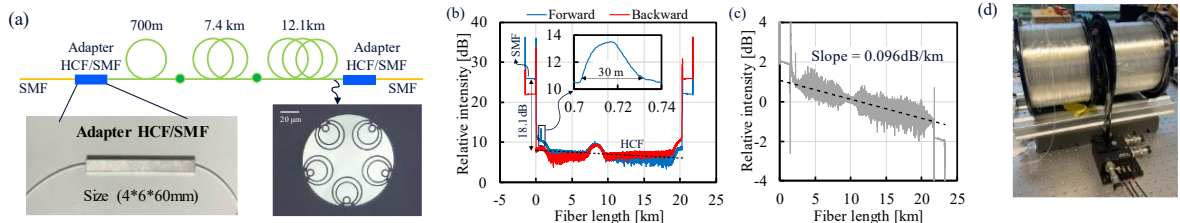

Figure 1: (a) The characteristics of the hollow core fiber are shown here (b) Standard OTDR trace of the HCF fiber acquired using few minutes of acquisition and with pulse duration ~ 100 ns. (c) The fit for the propagation losses using the difference between the forward and backward OTDR traces (d) Shows our HCF where we mount the spliced connection at 8.1kms on a translation stage to create acoustic excitations.

DFS has emerged as a powerful technology for monitoring acoustic events over long distances. These systems leverage homodyne detection with continuous probing and more recently using frequency sweeping and digital codes of type Golay to estimate the channel state (CS) [7]. As the signal is continuous, more energy can be sent in each probing interval (~ total fiber propagation time) allowing us to extract the CS more rapidly. Sensitivity and reach performances are limited by two independent effects in practical DFS implementations: additive noise from the optoelectronic components and coherence loss of the laser source. The former effect is magnified by the weak backscattered intensity level especially for HCF along with the speckle noise that randomly occurs along a fiber link. The latter effect can be alleviated by using stabilized laser source, here we use our custom designed compact Optical Frequency Discriminator to reduce the laser frequency noise in the spectral region of interest [8]. In addition, we use a combination of long probing codes at modulation rates reaching up to 500 Mbaud and use a matched filter at the receiver to alleviate SNR issues. Here we show how to develop a large dynamic range DFS system adapted for HCF systems.

In Figure 1d, we show the fiber spools of our HCF where the second splicing at 8.1 km is mounted on a 3-axis translation stage to create acoustic oscillations. In Figure 2a, we show our DFS setup where the retro-reflected signal from the sensed fiber is separated using an optical circulator and detected using coherent receiver. We multiplex a second wavelength carrying 1.2 Tbps of traffic using a commercial Nokia PSI-M transponder. The signal spectrum is shown in the Figure 2b with both sensing (in black box) and communication channel. At the output of the HCF, we demultiplex the two wavelengths, add arbitrary insertion loss using a variable attenuator and monitor the signal quality by measuring the uncorrected code blocks (UCB) i.e., the normalized amount of unsuccessful decoded blocks after the Forward Error Correction. In Figure 2c, we show the impact of increasing launched power on the signal quality. At launched powers below 13 dBm, we observe complete loss of traffic i.e., UCB of 1 as the power received by the PSI-M doesn't have sufficient optical signal to noise ratio for successful decoding due to added attenuation. Finally, we observe zero UCB with launched powers reaching up to 23 dBm per channel.

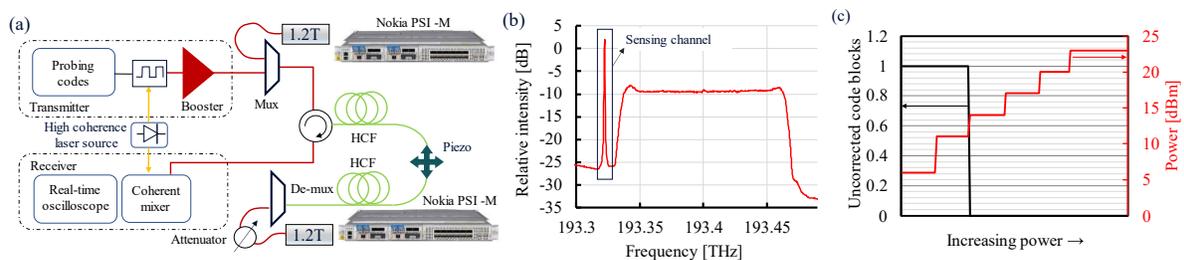

Figure 2: (a) High resolution coherent OTDR/DAS setup along with live traffic. (b) Spectrum of the transmitted signal showing the sensing channel (in black box) and the carrier next to it carrying live traffic. (b) Uncorrected code blocks at the traffic when the power at the input is varied showing no impact at high input powers.

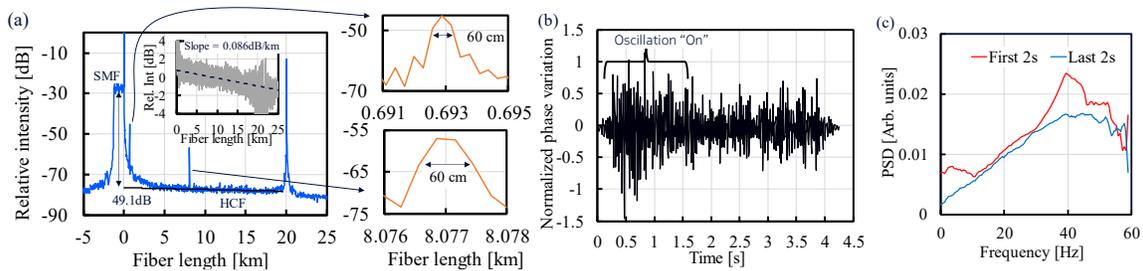

Figure 3: (a) High resolution coherent DFS measurement as a function of distance. We show the FWHM of the two spliced peaks on the right (b) Normalized phase variation as a function of time where we induce artificially an oscillation in first 2 seconds and turn it off. (c) We show the PSD of the signal in the first and last 2 seconds.

### 3. Experimental results

We use advanced probing code design derived from [7], achieving code length that can sense fiber lengths up to 600 kms, thanks to the stabilized laser source we are able to use code lengths reaching ~ few milliseconds and acquisition time of few seconds. The code is made of $2^{21}$ symbols at up to 500 Mbaud, leading to a 0.3 m native gauge length, and the permanent repetition of the code allows to capture mechanical events over a 120 Hz bandwidth. We first estimate the intensity of the retro-reflected light as a function of distance as shown in Figure 3a. We observe that the RBC from the SMF is about 49.1 dB higher than that from the HCF. The residual RBC from the fiber is estimated to be about -125.1 dB/m. We show in Figure 3a (right) the FWHM of the peaks around the two spliced regions revealing a value of 60 cm corresponding to our 0.3 m resolution. Moreover, in the Figure 3b we show the difference between the forward and backward measured DFS to estimate the propagation losses as a function of time leading to a value of 0.086 +/- 0.005 dB/km with 95% confidence interval consistent with the OTDR measurements. We measure an end-to-end loss of 2.3 dB, which suggests that the combined loss from the two adapters and all the splices is approximately 0.56 dB. Hence, the losses due to the MFA is below 0.3 dB per adapter, which aligns with the state-of-the-art ultra-low loss adapters proposed in [9]. Now, we perform acoustic sensing measurements by creating an oscillation on the spliced region. Due to practical constraints, we created the oscillation near the splicing point at 8.1 km as shown in Figure 1d. In Figure 3c, we show the measured relative phase as a function of time. In Figure 3d, we show the power spectral density of the phase in the first 2 seconds of acquisition and compare with the PSD of the last 2 seconds. We use a high pass filter in the phase estimation to remove ultra-low mechanical oscillations below 20 Hz, and hence the low noise at small frequencies, however we can clearly see the peak arising at 40 Hz oscillation at RBC powers below -100 dB/m from the spliced region despite the elevated RBC from the air of ~ -93 dB/m.

### 4. Conclusion

We showed the practical challenges of performing OTDR with hollow core fiber and demonstrated a sub-meter resolution OTDR using DFS setup and characterized an ultra-low loss anti-resonant HCF with propagation losses 0.086 +/- 0.005 dB/km. We then showed that we can achieve a low resolution of 0.3 m with few seconds of acquisition over 20 kms with precise splice localizations. Finally, we show for the first time acoustic sensing and accurately detected the phase variation on ultra-low loss HCF. These results will pave the way for high resolution monitoring of propagation losses and DFS for hollow core systems without having to compromise on the quality of the coupling between SMF and HCF and with no impact on the traffic flow even when it is present adjacent to the sensing channel.

### References
[1] Y. Chen, et al., "Hollow Core DNANF Optical Fiber with <0.11 dB/km Loss," in OFC 2024, paper Th4A.8.
[2] E. Numkam Fokoua, et al., "Theoretical analysis of backscattering in hollow-core antiresonant fibers" APL Photonics 2021; 6 (9): 096106.
[3] E. Ip et al., "First Field Demonstration of Hollow-Core Fibre Supporting Distributed Acoustic Sensing and DWDM Transmission," in ECOC, paper Th1F.1, 2024.
[4] X. Wei et al., "Distributed Characterization of Low-loss Hollow Core Fibers using EDFA-assisted Low-cost OTDR instrument," OFC 2023.
[5] R. Slavík et al., "Optical time domain backscattering of antiresonant hollow core fibers," Opt. Express 30, 31310-31321 (2022)
[6] V. Michaud-Belleau, et al., "Backscattering in antiresonant HCF: over 40 dB lower than in standard optical fibers," Optica 8, 216-219 (2021)
[7] C. Dorize et al., "From Coherent Systems Technology to …," IEEE JLT, vol. 41, no. 4, pp. 1054-1063, 15 Feb.15, 2023
[8] R. Boddeda et al., "Demonstration of MIMO-DFS over 100km of unamplified SSMF Link using Active Laser Drift Stabilization and Optimized Probing Codes" OFC, 2025
[9] D. Suslov et al., " Low loss and broadband low back-reflection interconnection between a hollow-core and standard single-mode fiber," Opt. Express 30, 37006-37014 (2022).